
\def\gtorder{\mathrel{\raise.3ex\hbox{$>$}\mkern-14mu
             \lower0.6ex\hbox{$\sim$}}}
\def\ltorder{\mathrel{\raise.3ex\hbox{$<$}\mkern-14mu
             \lower0.6ex\hbox{$\sim$}}}
\magnification=\magstep1
\line{\hfill Version of 30 June 1995}
\bigskip
\def\et{{\sl et al~}}
\font\BF=cmbx10 scaled \magstep1
\font\SM=cmr6 scaled \magstep1
\centerline{\BF STATISTICS OF N-BODY SIMULATIONS. III. UNEQUAL MASSES}
\bigskip
\centerline{Mirek Giersz\footnote{$^1$}{Now at the N.  Copernicus
Astronomical Centre, Bartycka 12, 00-716 Warsaw, Poland} and Douglas C.
Heggie}
\medskip
\centerline{University of Edinburgh,}
\centerline{Department of Mathematics and Statistics,}
\centerline{King's Buildings,}
\centerline{Edinburgh EH9 3JZ,}
\centerline{U.K.}
\bigskip

{\bf ABSTRACT}

{\parindent=0pt
We describe results from large numbers of $N$-body simulations
containing from $250$ to $1000$ stars each.  The distribution of stellar masses
is a power law, and the systems are isolated. While the collapse of the core
exhibits the expected segregation of different masses, we find that the
post-collapse evolution is, at a first approximation, homologous. This is quite
surprising because there is no reason for supposing that mass segregation
should
not continue to have a substantial effect on the evolution of the cluster. In
fact the spatial distribution of the mean stellar mass is nearly static
throughout the post-collapse regime, except for the overall expansion of the
systems, and this helps to explain why the post-collapse evolution is nearly
self-similar. Self-similarity is also exhibited by the distribution of
anisotropy and the profile of departures from equipartition, which show little
change during the post-collapse phase. The departures from energy equipartition
and isotropy are small in the core and increase with radius. During
post-collapse evolution massive stars (mainly) are removed from the system by
binary activity. This effect dominates the preferential escape of low-mass
stars
due to standard two-body relaxation processes.}

\bigskip


\bigskip
\bigskip

{\bf 1  \enskip INTRODUCTION}
\medskip

This is the third in a series of papers in which we investigate the evolution
of
stellar systems using $N$-body simulations.  This has often been done before,
but one of the main drawbacks of most such studies is the poor statistical
quality of much of the data.  Even in a system with 1000 stars, the sampling
noise in the results becomes overwhelming whenever measurements are taken from
a
small subset of the stars.  Therefore it is difficult to carry out detailed
quantitative studies of many interesting phenomena, e.g.  the evolution of the
core, the spatial distribution of anisotropy, the escape rate, and so on.  This
problem becomes much more acute whenever stars of different masses are present,
if it is desired to investigate mass segregation, departures from
equipartition,
the differential escape rate, etc.  Indeed this paper is concerned with
precisely this kind of issue.

The main technical improvement in the present research is an extremely simple
one.  Instead of presenting results from one simulation, or only a small
number,
we combine statistically the results from several dozen simulations.  The
intention is to improve the signal-to-noise of our results to the point at
which
reliable quantitative data can be presented, and compared with the results of
simplified models (e.g.  gaseous models, or Fokker-Planck models.) An
alternative method of improving the statistical quality would be to simulate a
single, larger system.  An advantage of that approach is that it works in the
direction of greater realism (if the intention is to understand the dynamical
evolution of globular star clusters).  But a much greater increase in
computational effort is required for a given improvement in signal-to-noise,
compared with our strategy of simulating large numbers of more modest
simulations (see Giersz \& Heggie 1994a).  This strategy is especially
effective
now that parallel computers have become readily available.

The present paper differs from the first two papers in this series (Giersz \&
Heggie 1994a, 1994b, hereafter referred to as Papers I and II) by dealing with
the case of unequal stellar masses.  This greatly extends the range of
parameters to be explored, but we have restricted ourselves to parameters
comparable to those in the well known Fokker-Planck survey carried out by
Chernoff \& Weinberg (1990).  (One of the motivations for our research was the
desire to test their results using $N$-body methods, which make fewer
assumptions.) On the other hand we still consider systems which are isolated
(without tidal effects) and in which the stellar masses are constant, i.e.  we
do not model systems in which the stars lose mass through internal evolution.
Those two processes are included in later sets of models which we discuss in
two
further papers.

The plan of the paper is as follows.  We begin with a discussion of the initial
parameters which we adopted, such as the slope and range of the stellar mass
function.  Then subsequent sections discuss, in turn, the spatial evolution of
the systems, and in particular the segregation of masses.  Next we discuss the
distribution of velocities, and in particular the extent to which equipartition
is obeyed in the systems at different radii.  Then there follow sections on the
evolution of the core and of the binaries which mostly form there.  Towards the
end of the paper we discuss global parameters, e.g. the evolution of the total
mass and energy, with a detailed description of the differential rate of escape
(i.e. its dependence on stellar mass).  The last section discusses one or two
points of interest and sums up.

\bigskip

{\bf 2 \enskip ORGANISATION OF THE CALCULATIONS}
\medskip

As in earlier papers of this series, all models reported here started as
Plummer
models.  The initial mass spectrum was chosen to be a power law of index $2.5$,
i.e. $dN(m) \propto m^{-2.5}dm$, in a range such that $m_{max}/m_{min} = 37.5$.
This choice was made in order to facilitate comparison (in later papers in this
series) with results of Chernoff \& Weinberg (1990), who used a very similar
spectrum.

Though the spectrum of masses is continuous, for purposes of analysis the
masses
were grouped into 10 classes, along the lines of those used by Chernoff \&
Weinberg (see Table 1 in this paper).  (In their Fokker-Planck calculations it
is impractical to include a continuous spectrum of masses.  Therefore a finite
number of masses are used, and after considerable experimentation Chernoff \&
Weinberg found that use of 10 masses yielded results quite similar to those
using larger numbers of masses.)

The calculations were carried out on the Edinburgh Computing Surface (ECS),
which was a transputer array containing about $400$ processors (cf. Paper I).
The code used was NBODY5 (Aarseth 1985).  Three sets of calculations were
carried out, in which the number of stars was $N = 250$, $500$ and $1000$,
respectively.  In each set, the respective number of models was $56$, $56$ and
$40$.  Virtually all runs were completed to time $t = 400$.

The organisation of the output was handled somewhat differently than in earlier
papers in this series.  The reason for this was fairly directly related to the
use of a spectrum of masses, divided into the 10 classes described above.
Because of the steepness of the mass spectrum, the average number of stars in
the class with heaviest stellar masses was less then one for $N = 250$.  Since
we wished to measure Lagrangian radii for each mass group in a uniform manner,
and consistently with earlier papers where data for the inner 1\% Lagrangian
radii were discussed, it was necessary to combine data from all models at the
same $N$-body time and also to combine data from three mass bins corresponding
to the most massive stars.  (Even so, with only $56$ models in each series,
there are still only of order $150$ stars in the heaviest mass group).  Since
the calculations produce data at different rates, however, this approach meant
that it was necessary to store data on each star from each model at each output
time.  In order to conserve space this was done by coding data in such a way as
to allow it to be reconstructed to adequate accuracy in subsequent analysis.

The data stored (to limited accuracy) for each star was as
follows:

{\parindent=0pt
1.  mass;

2.  distance from the density centre (defined as in Casertano \& Hut 1985, with
a modification for the presence of a mass spectrum, described below); and

3.  radial and tangential velocity (the radial direction being that from the
density centre).

In addition, the following details of all regularised binaries, and all
unregularised binaries with binding energy $\geq 0.1 kT$ (where the current
mean
kinetic energy of all single stars is $3kT/2$) were output:

4.  masses of components;

5.  separation of components, and distance of binary centre of mass from the
density centre;

6.  internal energy of the binary and its eccentricity;

7.  radius of neighbour sphere and number of neighbours (defined as in NBODY5);

8.  names of binary components.}

The typical size of an output file from $40$ runs at $N = 1000$ for $400$
output
times was of the order of $125$Mb.  Analysis of such a file proved to be
manageable, if awkward, even on a diskless workstation requiring access to
files
over a fairly busy ethernet.

In addition, each model outputted some preanalysed data, as described for
equal-mass models in Paper I.  For example, the values of ten Lagrangian radii
(defined below) were output.  As in our earlier work, the different sets of
calculations did not all output every item in the list in the appendix to Paper
I: the data selected for output was changed slightly as our interests and
experience developed.

The previous discussion refers to ``Lagrangian radii'', a term which requires
somewhat careful definition in the context of these multi-mass models.  First,
these radii are referred to the ``density centre'', which is defined as
follows:
for each star the density is defined to be proportional to $M_5r_5^{-3}$, where
$r_5$ is the distance to the fifth nearest neighbour (cf.  Casertano \& Hut
1985) and $M_5$ is the total mass of the five nearest neighbours.  Then the
density centre is defined as in Aarseth \& Heggie (1992).  Incidentally the
density centres of different models were found to gradually move away from the
origin (whose position and velocity coincide initially with those of the
barycentre) in different directions.  The root mean square value of each
coordinate of the density centre (averaged over all systems in a given series)
increases nearly linearly with time in the post-collapse regime, roughly as
$0.05t$ for $50 \leq t\leq 200$ for $N=250$.

With the density centre thus defined, a Lagrangian radius is defined to be the
radius of a sphere centred on the density centre and containing a fixed
fraction
of the total bound mass of the system.  (When this does not correspond to a
whole number of stars, interpolation of ordered radii was used as in Paper I.)
Because of mass segregation and the presence of a spectrum of masses, it can
happen that the innermost radius, which corresponds to a mass fraction of
$1\%$,
may contain less than one star in a single model.  For this reason the
Lagrangian radii discussed in this paper were actually determined differently.
For each output time, the radii of all stars (measured from the density centre
of the model to which it belongs) were aggregated from all models in a given
series, and then Lagrangian radii were determined as above from this much
larger
set of stellar data.  It was hoped that this would improve the determination of
the inner Lagrangian radii, and avoid the difficulties associated with mass
segregation, as described above.

\bigskip
{\bf 3 \enskip SPATIAL EVOLUTION}
\medskip
{\bf 3.1 \enskip The entire mass distribution}
\smallskip

The essential facts about the evolution of the mass distribution are apparent
from Fig.1.  The collapse of the core is much more rapid than in the equal mass
cases considered in Papers I and II (cf. Table 2).  Indeed the ratio of
$t_{coll}$ (between the values for unequal and equal masses) is a decreasing
function of $N$ even for the largest value of $N$ that we have studied.  At
first sight this appears to contradict the result that would be expected from
the standard theory of relaxation, which is that this ratio should be
independent of $N$.  However, it was argued long ago by H\'enon (1975) that the
coefficient $\gamma$, which appears in expressions for the relaxation time in
the Coulomb logarith $\ln\gamma N$, is generally smaller in systems with
unequal
masses than in those with equal masses, by as much as a factor of 5.  Indeed
his
theory yields the result $\gamma \simeq 0.044$ for the mass function used in
our
models.

Now we turn to the numerical data (Table 2).  If it is assumed that the
collapse
time (as a function of $N$ for unequal masses) is proportional to the initial
half-mass relaxation time, then we find that the results of Table 2 are roughly
consistent if $\gamma\sim 0.015$, which is a factor $7$ smaller than the value
found in Paper I for systems with equal masses.

An independent method of determining $\gamma$ is comparison of $t_{coll}$ with
the results of a Fokker-Planck computation with the same initial conditions and
mass spectrum.  This has been done using an isotropic code (Inagaki \& Wiyanto
1984), from which the value of $t_{coll}$ was obtained by finding the time at
which the central potential reached the minimum attained in the $N$-body
results
(Fig.12).  Scaling to the values of $t_{coll}$ in Table 2 yielded values of
$\gamma$ in the range $0.016\ltorder\gamma\ltorder0.026$.

Whatever the correct value of $\gamma$, we have seen that there are three
independent lines of evidence for a substantially smaller value than in the
case
of equal masses.  This implies that considerable care is required if the
results
of models with small $N$ are to be scaled to much larger systems.  For example,
when $N = 1000$ the relaxation times estimated by H\'enon's value and the value
$\gamma= 0.02$ differ by about 25\%.  Even the largest of our models are,
therefore, a somewhat uncertain guide to the absolute time-dependence of
relaxation processes in larger $N$-body systems.

The post-collapse evolution is, to a first approximation, self-similar, as one
can see from the fact that the curves in Fig.1 are very nearly parallel, in the
sense that all can be brought nearly into coincidence by a suitable vertical
displacement.  This is evidently not true in the early post-collapse phase for
the outermost radii (mass fractions of $75\%$ or more).  These aspects of the
evolution are illustrated qualitatively in Fig.2, which shows the surface
density profile scaled (radially) to the half-mass radius.  In post-collapse
the
profiles could be brought into near-coincidence by a vertical shift.
Nevertheless even for the latest profiles shown in Fig.2 the outermost radius
at
which the density is plotted (which corresponds to a mass fraction of $90\%$)
moves outwards slowly as time advances, indicating a modest but sustained
departure from homologous evolution.

Though there are departures from self-similarity, the half-mass radius still
varies in approximate accord with naive theoretical expectations,
i.e. $r_h\propto t^{2/3}$.  In fact there are small departures from this
formula. To clarify this point we carried out a similar exercise to that
described in Paper II, i.e. we fitted the following functions to the $N$-body
data: $$\eqalign{ N &= a(t-t_0)^{-\nu}, \cr r_h &= b(t-t_0)^{(2+\nu)/3},\cr
(Nr_h^3/\bar m)^{1/2} &= c(t-t_0),\cr
}\eqno(1)$$
where $N$ is the number of bound stars, $r_h$ is the half-mass radius,
$t$ is time, $\bar m$ is the mean individual mass of stars inside $r_h$,
and $a$, $b$, $c$, $\nu$ and $t_0$ are fitting parameters.
The values obtained are given in Table 3.

According to standard relaxation theory (Spitzer 1987) the coefficient $c$
should be proportional to the Coulomb Logarithm, usually taken as $\ln(\gamma
N)$ for some value of the constant $\gamma$. But as can be seen from results
presented in Table 3 it is a good approximation to assume that $c$ is constant,
i.e. independent of $N$, and this is inconsistent with any plausible value of
$\gamma$.  A similar conclusion follows from consideration of the values of
$b$.
We have no definitive explanation for these results, though it should not be
concluded that the evolution of systems of stars with unequal masses behaves in
a manner inconsistent with standard relaxation theory.  At some level of detail
the mass distribution evolves differently in systems with different $N$, and
also our results have been obtained by fitting to the evolution over an
$N$-dependent number of relaxation times; for example, the factor by which
$r_h$
expanded was very different in the different models.  These $N$-dependent
effects may have conspired to mask the expected $N$-dependence of the Coulomb
logarithm.

It is to be noted that Fig.1 illustrates {\sl projected} Lagrangian radii, i.e.
containing a given fraction of the projected mass.  The corresponding picture
for the conventional (three-dimensional) Lagrangian radii is qualitatively
similar, at least in respect of the aspects to which attention has been drawn
above.  The three-dimensional half-mass radius is larger in the post-collapse
phase by about 30\%.  In the same way the space density profiles exhibit very
similar features to the surface density profiles shown in Fig.2.  Also the
results for $N=1000$ and $250$ do not differ qualitatively from those shown
here
for $N=500$. The main quantitative difference, which is the time scale of the
evolution, has already been described above.

\medskip
{\bf 3.2 \enskip Mass segregation}
\medskip

At first sight it is surprising that the structure of the cluster in
post-collapse evolution appears to be so nearly self-similar, and that the
naive
theoretical expectation, eq.(1), is so well satisfied, because there is no
reason for supposing that mass segregation should have a negligible effect
after
core collapse.  Therefore, even though it remains true that the time scale for
post collapse expansion should be governed by a time of relaxation, it is not
as
clear as it is in the case of equal masses that this time scale may be
characterised by the half-mass relaxation time: as conventionally defined
(Spitzer 1987) this depends on the mean mass as well as the total mass and the
half-mass radius, and one might have expected the somewhat different time scale
for mass segregation to complicate the evolution of the half-mass radius.  In
the present section we shall discuss mass segregation, and we shall find an
empirical explanation for the fact that the evolution of the half-mass radius
is
so simple.  What is still lacking, however, is a theoretical understanding of
why mass segregation in post-collapse evolution takes the simple form observed.

The basic result is illustrated in Fig.3.  Extremely rapid segregation of
masses
takes place throughout core collapse.  The remarkable conclusion to be drawn
from this figure, however, is that the profile of mean masses evolves
remarkably
little throughout the remainder of the calculations, which amounts altogether
to
about 20 collapse times.  It is presumably this fact, i.e.  the fact that the
distribution of stellar masses is nearly static (in Lagrangian coordinates),
which helps to explain why mass segregation does not, as one might expect,
complicate the post-collapse evolution of the cluster.  Note, however, that
this
merely places the problem one step back: we have no explanation for the fact
that the distribution of masses at each Lagrangian radius shows such little
evolution after the initial phase of mass segregation.

While Fig.3 illustrates the projected distribution of the mean mass, the
spatial
distribution shows very similar features.  There is, however, a slight but
noticeable drop in the mean mass inside the 20\% radius in the early
post-collapse regime and for all Lagrangian radii during the advanced
post-collapse phase.  This drop is more pronounced than the corresponding drop
in the projected data (Fig.3), but still no greater than about 0.1 dex.  It is
not clear that this is correctly attributed to mass segregation, since stars of
high mass are removed from the core and the whole system by binary formation,
evolution and escape (\S6).

Since the distribution of masses is almost constant in the post-collapse
regime,
except for the overall expansion of the radii and the corresponding decrease in
the density, we illustrate the density distribution of the different masses,
binned as stated in \S2, at one point in the post-collapse phase (Fig.4).
Initially, of course, all density profiles would be identical, except for an
overall vertical displacement from one species to another. As can be expected
at
this point in the post-collapse evolution, however, the effects of mass
segregation are clearly visible.  The most massive stars dominate the central
parts of the systems and the less massive stars dominate the halo. As a rule,
the larger the stellar mass the greater the concentration to the centre, with
one exception: stars from the most massive bin are not, as one might have
expected, confined to the central and middle parts of the system; they are
present in the outer halo as well, at distances even greater than those of the
least massive stars. (More precisely, their 90\% Lagrangian radius exceeds that
of the least massive stars.)  This behaviour is probably connected with the
ejection of massive stars and binaries from the core to the outer parts of the
system during interactions between binaries and field stars.  This is evidence
that binary activity can disturb the otherwise clear picture of mass
segregation
in these systems.

\bigskip
{\bf 4 \enskip EVOLUTION OF THE VELOCITY DISTRIBUTION}
\medskip
{\bf 4.1 \enskip The overall velocity dispersion}
\medskip

Many aspects of the overall evolution of the velocity distribution are
reflected
in Fig.5.  After the relatively brief period of core collapse the decline in
the
mass-weighted root mean square projected speed is nearly homologous.  The
profiles of the projected velocity dispersion behave in a manner very similar
to
that of the density profiles (cf.  Fig.2): if the radial coordinate is scaled
by
the projected half-mass radius the profiles at different times differ from each
other only by a vertical shift in a logarithmic plot, to good approximation.
This is also true of the three-dimensional velocity dispersion profiles, though
they are very slightly steeper: between the centre and the $90\%$ Lagrangian
radius the difference in rms speed is about $0.36$ dex for the
three-dimensional
profile, and about $0.33$ for the projected profile.

\medskip
{\bf 4.2 \enskip Equipartition}
\medskip

The most interesting kinematic distinction between the models discussed here
and
those analyzed in Papers I and II concerns the question of equipartition, or
more generally how the velocity dispersion differs from one population to
another.  Just as we have attempted to summarise the distribution of masses by
the mean mass profile (Fig.3) we may begin by attempting to define a single
measure of the extent to which different masses are in equipartition, as
follows.

For a given Lagrangian shell let $T_k$ be the mean kinetic energy of the stars
in that shell in mass class $k$, with $1\le k\le 10$ (\S2), and $T$ the mean
kinetic energy of all stars in the shell.  If there happen to be no stars in a
given mass class in that shell then we define $T_k = 0$.  Next we define the
`equipartition parameter' as $$ B = {\sum_k T_k\over N_>T}, \eqno(2) $$ where
$N_>$ is the number of mass classes for which $T_k>0$.  The obvious feature of
$B$ is that in the case of equipartition (of kinetic energies) we should have
$T_k = T$ for all $k$, and so $B = 1$.  Likewise, in the case of equipartition
of velocities we would have $B = \sum m_k/(10\langle m\rangle)$, where $m_k$ is
the mean mass of stars in the $k$th mass class.  Notice that $\sum
m_k/(10\langle m\rangle) \not= 1$, since the average on the left hand side is
not weighted by the number of stars in each mass class.  For our initial mass
function, for example, we would have $B = 3.8$ (i.e. $\log B = 0.58$) in the
case of equipartition of velocities.

Clearly one may compute $B$ either for three-dimensional Lagrangian shells or
for projected annuli, in which case $T_k$ could be defined as the mean kinetic
energy of motion in the line of sight.  This has been adopted for Fig.6, which
shows profiles at various times.  The curves are noisier than those previously
shown, because the small number of stars of high mass have relatively high
weight in the numerator.  Nevertheless one can easily observe an extremely
rapid
evolution (in the first $5$ time units for $N=500$) from the initial flat
profile to one which remains almost constant throughout the remainder of core
collapse and in the post-collapse phase provided (as has been done here) the
radial coordinate is scaled by the projected half-mass radius.  In the core the
value is close to that expected for equipartition of energy, whereas in the
outermost parts of the system the departure from energy equipartition, to the
extent that this is measured by $B$, is even greater than it was initially.

More detail is shown in Fig.7, which gives the profiles of the mean square
projected velocity as a function of projected radius, for all mass groups.  (We
give this for only one time, since Fig.6 suggests that there is little
evolution
after an initial short period of adjustment.)  Unfortunately, because of the
small number of stars in each mass bin (particularly the massive bins) the data
presented in Fig.7 are very noisy. However, the main features of this picture
are clearly visible. In the core, dominated by massive stars, the mean square
projected velocities follow closely the rule of energy equipartition. In the
outer halo, dominated by less massive stars, the mean square projected
velocities behave differently: they are nearly the same for most of the mass
bins. This reflects the initial conditions, for which all mass bins were in
velocity equipartition. In the outer halo the relaxation time is very long
(compared to that in the core, or even at the half-mass radius), and so the
evolution is very slow and the initial conditions are well preserved.
Therefore
the equipartition parameter $B$ increases towards the outer parts of the system
(see Fig.6).

For the most massive bin the velocity is nearly constant throughout the system.
The explanation for this may be connected again with binary activity, which
removes stars and binaries from the core and ejects them into the halo with
velocities high compared with most other stars in the halo. Because massive
stars are preferentially removed from the core the behaviour of the most
massive
bin is markedly different from that of the others.

\medskip
{\bf 4.3 \enskip Anisotropy}
\medskip

In discussing the anisotropy we shall concentrate on projected values (Fig.8).
Thus each annulus is bounded by circles containing fixed fractions of the total
mass (relative to the density centre of each model), and within each annulus we
denote by $\langle v_r^2\rangle$ and $\langle v_t^2\rangle$, respectively, the
mean square radial and transverse components of the projected velocity.  Then
the anisotropy, defined by $\langle v_t^2\rangle/\langle v_r^2\rangle$, is
plotted at a radius given by the mean of the radii of the bounding circles
scaled by the projected half-mass radius.

It can be seen from Fig.8 that the evolution of the anisotropy profile occurs
on
a longer timescale than that of mass segregation (Fig.3) and the tendency to
equipartition (Fig.6). (This difference of timescales is not really surprising,
since the main evolution of the anisotropy occurs at large radii, unlike the
other two phenomena.) When $N = 500$ the profile continues to evolve in the
outer parts until about $t \sim100$, and only thereafter remains relatively
constant (when the radius is scaled by the expanding projected half-mass
radius).

As already stated the values plotted in Fig.8 are projected values.  In the
quasi-steady phase after $t\sim100$ the three-dimensional values of
$\log(\langle v_t^2\rangle/\langle v_r^2\rangle)$ are about $-0.6$ at the
outermost radii shown on Fig.8 (i.e.  at a similar value of the
three-dimensional radius scaled by the three-dimensional half-mass radius).
Though this might be taken to imply that the three-dimensional anisotropy is
smaller, it must be recalled that the value of the logarithm in an isotropic
distribution is $\log2$ in three dimensions, and $0$ for projected data.

Another point to be made about these data concerns the $N$-dependence of the
aniso\-tropy.  As in the case of equal masses (Giersz \& Spurzem 1994) the
anisotropy in the outer parts saturates at a value which is greater (i.e.  more
radially anisotropic) for smaller $N$.  For $N = 250, 500$ and $1000$ the
values
of the logarithm (three-dimensional anisotropy) are about $-0.85,~-0.63$ and
$-0.48$, respectively.  This may have a similar explanation to that of the
faster expansion of the outer Lagrangian radii for smaller $N$ in systems with
equal masses (cf. Paper II), i.e. the fact that the relatively shallower
potential well of systems with small $N$ makes it easier for stars which are
ejected from the core to populate the far halo.

The way in which the anisotropy is distributed among the different mass classes
is illustrated in Fig.9, which exhibits the projected anisotropy profile for $N
= 500$ at $t=300$, i.e. well into the phase in which the overall profile
appears
to have settled down into equilibrium (as a function of radius scaled by the
half-mass radius). Generally, the core is isotropic for all mass bins (with
very
substantial fluctuations for massive bins). In the outer halo the anisotropy is
bigger for the bins corresponding to the most massive stars; as with
equipartition and the velocity dispersion, the most massive bin is exceptional.
This is again consistent with the suggestion that the massive stars are ejected
from the core into the halo by binary activity.

\bigskip
{\bf 5 \enskip EVOLUTION OF THE CORE}
\medskip

In systems of stars of equal mass there are already two commonly used measures
of the core radius.  One, stemming from the work of King (1966), defines the
core radius $r_c$ in terms of the central one-dimensional velocity dispersion
$\sigma_c$ and the central mass-density $\rho_c$ by the formula $$ r_c^2 = {{9
\sigma_c^2}\over {4\pi G\rho_c}}.\eqno(3) $$ The other, derived from the work
of
Casertano \& Hut (1985), defines it as $$ r_c^2 = {\sum \rho_i^2 r_i^2\over\sum
\rho_i^2},\eqno(4) $$ where $\rho_i$ is a measure of the density at the
location
of the $i$th star, and $r_i$ is its distance from the density centre (\S2).
(The formula proposed by Casertano \& Hut actually differed from eq.(4) in that
all the terms were unsquared.) In our work we have estimated $\sigma_c$ and
$\rho_c$ from the innermost 1\% by mass (cf. below), while $\rho_i$ was
estimated as in \S2.  It is known (cf. Aarseth \& Heggie 1992) that the
definition in eq.(3) gives a larger value than eq.(4) when applied to a Plummer
model, which is the starting configuration of our calculations.  But in our
unequal-mass models it quickly becomes the smaller, exhibiting a much deeper
collapse: for $N = 500$, for instance, the core radius defined by eq.(3)
decreases by a factor of $3$ approximately, whereas the decrease for $r_c$
defined by eq.(4) is only by a factor of about $1.4$.  Throughout the
post-collapse evolution eq.(4) gives a core radius which is larger than that
defined by eq.(3) by a fairly consistent factor which is nearly $2$.  We prefer
eq.(3), entirely on the grounds that it has a dynamical significance, which has
not been demonstrated for eq.(4).

As for the value of $r_c$ itself, the simplest way of expressing the result for
post-collapse is to give the ratio to the half-mass radius (Table 2), which is
almost constant in this phase of the evolution.  These results are quite
interesting, as they contradict the theoretical expectation for equal masses,
which implies that, for large $N$, the ratio should vary as $N^{-2/3}$ (Goodman
1987).  Whether this implies that the models have not yet reached the
asymptotic
behaviour of systems of sufficiently large $N$, that systems with stars of
unequal mass behave in a qualitatively different way, or that our adopted
definition of core radius is faulty in some way for systems with unequal
masses,
cannot be decided on the basis of our data.  Certainly, our conclusion for
systems of equal mass (Paper II) was that the size of the core was roughly in
line with theoretical expectations, even for the small values of $N$ that were
studied.

The definition of core radius having been decided on, we can readily determine
the number of stars within this radius (Table 2), from which it appears that
$N_c\simeq N/60$ throughout almost all of the post-collapse evolution.  The
mass
of stars within $r_c$ is approximately $0.05$, independently of $N$, and so the
mean stellar mass in the core is $3/N$ approximately, i.e. about three times
the
initial global mean mass (cf. also Fig.3, which gives a similar value for the
{\sl projected} central mean mass).

Our estimate for the central velocity dispersion, which we now discuss, is
obtained from the mass-weighted mean square speed of stars within the innermost
$1\%$ by mass.  Even though we include all stars from all models (at given
$N$),
because of mass segregation the stars in this zone are relatively massive and
few in number.  Therefore the estimate of the mean square speed is subject to
quite large fluctuations, especially in relation to the rather small changes in
this quantity during the evolution.  Fig.10 shows the results for all values of
$N$ that we have studied.  One reason why these results are interesting is the
initial decline in the early phase of core collapse, though because of the
fluctuations the result is convincing really only for $N = 1000$.  A similar
drop was noticed in the equal-mass calculations discussed in Paper I, but it is
more natural here to interpret it as a consequence of mass segregation and the
tendency towards equipartition.  A similar drop can be seen as well in data
presented by Chernoff \& Weinberg (1990) for Fokker-Planck calculations for
King
models with unequal masses.  As already stated, and discussed further below,
the
core quickly becomes dominated by massive stars, whose rms speed drops below
that of the stars of lower mass.  Thereafter the core collapse, which involves
mainly the dominant massive stars, leads to the expected increase in the rms
speed.

The fact that the central density rapidly becomes dominated by the most massive
stars is responsible for the evolution of the mean mass within the innermost
Lagrangian radius, as shown in Fig.3.  Fig.11 gives more detail, showing the
contribution to the central projected density from the several mass bins.
(Note, however, that the density is influenced also by the sizes of the mass
groups, which are uniform in $\log m$.)

Another core parameter of importance (e.g.  for escape of binaries, and the
efficiency of heating by three-body binaries) is the depth of the potential
well
at the centre of the system (Fig.12).  The minimum is less sharp than in the
case of equal masses (Paper II), perhaps because of an even greater relative
spread in the times of core collapse in the models with unequal masses, and
perhaps because the presence of a few massive stars in the core causes larger
fluctuations than in the case of equal masses. Note that the minimum depth of
the potential is increasing with $N$.  This result is quite interesting as it
contradicts theoretical expectation (e.g. Goodman 1987) and numerical results
for equal masses (Paper II), which implies that the minimum of the potential
decreases with increasing $N$.  We interpret this as a richness effect: the
maximum individual stellar mass which is present in a given small $N$-body
system may be much less than the maximum allowed mass, if the mass spectrum is
steep.  (There is evidence for this in Table 1, which shows that the fraction
of
mass in the heaviest mass group is an increasing function of $N$.)  One
consequence of this is that a small $N$-body system is unlikely to be able to
form a binary from stars close to the maximum allowed mass, and therefore
binary
heating is likely to be less efficient than in a system with sufficiently large
$N$.

All the results on core evolution displayed so far concern the time dependence,
but it is also instructive to display the evolution of two fundamental core
parameters against each other.  For Fig.13 we have chosen the central
mass-density and the central mass-weighted root mean square speed.  This bears
some resemblance to the corresponding result for equal masses (Paper II).  The
dip in the central velocity dispersion during core collapse has already been
discussed, but the post-collapse behaviour is another example of the
surprisingly simple behaviour of these unequal-mass models in this phase of the
evolution: if it is assumed that $\rho_c$ and $v_c^2$ are proportional to
values
at the half-mass radius, then it follows that $\rho_c v_c^{-6}\propto M^{-2}$.
Since the total mass $M$ varies slowly (cf. \S7), this predicts an approximate
power-law relation which is quite closely verified in Fig.13.  It is quite
surprising that the presence of a spectrum of masses does not invalidate such
simple arguments.
\bigskip

{\bf 6 \enskip BINARIES}
\medskip
{\bf 6.1 \enskip Definitions}
\medskip

For systems of stars with equal masses the notions of ``hard" and ``soft"
binaries are relatively uncomplicated.  When there is a spectrum of masses,
however, it becomes less clear what definition to adopt.  The simplest approach
would be to classify binaries according to the ratio $\varepsilon/(kT)$, where
$\varepsilon$ is the (internal) binding of the binary, and $1.5kT$ is the mean
kinetic energy of all stars in the neighbourhood of the binary.  On the other
hand it is not clear that this is the most useful definition when the masses
may
be very unequal.  In the case of equal masses the notions of ``soft" and
``hard"
are important usually because they correspond, roughly, to the distinction
between binaries which tend to be destroyed by encounters and those which
become
harder (Gurevich \& Levin 1950).  While work by Hills (1990) and by Sigurdsson
\& Phinney (1993) gives much useful information on the effect of encounters
between a binary and a single star when the masses are different, it is still
difficult to decide whether a binary, with a given energy and component masses,
will tend to harden or disrupt under the action of encounters with a field of
single stars, when the spectrum of stellar masses may be very time- and
position-dependent.

In view of this difficulty we have adopted a definition which is much simpler
to
implement, but is more difficult to interpret physically.  In what follows we
define a ``hard" binary to be a regularised binary, in the sense of the code
NBODY5 (Aarseth 1985).  We shall also refer more briefly to ``soft" binaries,
which are defined to be unregularised binaries with binding energies greater
than $0.1kT$ (cf. \S2).

Clearly, many binaries fall into neither of the classes which we term hard and
soft, because there will in general be a number of binaries whose internal
binding energies are too low to satisfy the criteria for regularisation.  There
is no evidence, however, for the presence of long-lived binaries which do not
fall into the class of regularised pairs.

\medskip
{\bf 6.2 \enskip Statistics of Binaries}
\medskip

As with equal masses (Paper II) the onset of the formation of hard binaries is
quite abrupt (Fig.14).  Though this is to be expected, being associated with
the
end of core collapse, what is perhaps more surprising is the result in the
post-collapse phase, where the number of hard binaries is nearly constant in
time, and independent of $N$.  In this phase the mean value is about 1.3.  For
systems of equal masses (Paper II) the corresponding result is about 2.4 (but
there are some indications for $N$-dependence of the form: $N_b \propto
N^{1/5.6}$).

The spatial distribution of these binaries is heavily centrally concentrated:
in
the early post-collapse phase about 80\% of the hard binaries lie within the
radius containing 20\% of the mass of all stars.  The central concentration of
hard binaries decreases somewhat in later phases of post-collapse expansion,
reaching a nearly constant value (as measured by the number of binaries within
the 20\% Lagrangian radius) for times greater than about $100$ $N$-body time
units (which apparently coincides roughly with the time at which the anisotropy
levels off). This drop is accompanied by an increase in the number of hard
binaries in the outer parts of the cluster. This is particularly clearly
visible
in data for $N = 250$, which extends to a larger number of collapse times than
our data for the larger systems.

The total internal energy of all bound hard pairs also shows the expected sharp
change towards the close of core collapse (Fig.15), and in the early
post-collapse phase it follows closely a power-law dependence, which can be
interpreted very roughly as follows.  It is known from the behaviour of the
half-mass radius (\S 3.1 above) that the ``external" energy of the bound stars
(i.e. excluding the internal energy of bound binaries) varies roughly as
$E_{ext}^b\simeq -C(t-t_0)^{-2/3}$, where $C$ is a positive constant, though
this form of dependence would be modified slightly if account were taken of
mass-loss.  If no binaries escape and we neglect the flux of energy associated
with escapers, it follows that the internal energy of bound binaries varies
roughly as $$ E^b_{int}\simeq E_0 + C(t-t_0)^{-2/3},\eqno(5) $$ where $E_0$ is
the initial energy of the system, i.e. $-0.25$ in our units.

In fact eq.(5) provides a reasonable fit to the data early in the post-collapse
phase, until one of the above assumptions begins to break down seriously.
Indeed the rise in $E^b_{int}$ after this early phase indicates the escape of
hard binaries (cf.\S7), and then the derivation of eq.(5) breaks down.  Before
this happens, however, we see from Fig.15 that $E^b_{int}$ falls below the
minimum value predicted by eq.(5), and the reason for this is that escape by
two-body encounters (\S7) tends to decrease the external energy of the system,
i.e. to make it more bound, thus decreasing the effective value of $E_0$.

The mean mass of the hard binaries is also instructive (Fig.16).  In the
innermost parts of the cluster they tend to consist of the most massive stars,
whose mean mass declines slowly with time as the most energetic (and most
massive) binaries escape or are ejected to the outer parts of the cluster
(\S7).
Indeed this gives rise to the perhaps rather surprising behaviour of the hard
binaries in the {\sl outer} parts, whose mean mass tends, if anything, to
increase.  It is in the intermediate radii that the mean masses of binaries are
smaller, and indeed decline relatively rapidly as the evolution proceeds.

The assertion that the massive binaries in the outer parts are ejected from the
core gains further support from an examination of the eccentricities of the
hard
binaries.  Outside the half-mass radius the mean eccentricity of hard binaries,
which is about 0.8, significantly exceeds that of the hard binaries inside
$r_h$
(about 0.7, i.e. close to the thermal average value of $2/3$ (Jeans 1929)).

\bigskip

{\bf 7 \enskip GLOBAL EVOLUTION}
\medskip

{\bf 7.1 \enskip Escape}
\medskip

In the previous section attention was already drawn to the behaviour of massive
hard binaries, which tend to be ejected into the outer parts of the cluster if
not out of the cluster altogether.  Surprisingly, perhaps, this escape
mechanism
is sufficiently effective that the mean mass of the remaining bound members of
the cluster tends to {\sl decrease}, whereas one might have thought that the
preferential escape of low-mass stars in two-body encounters (H\'enon 1969)
would lead to an increase in the mean mass.  This is illustrated in Fig.17,
which shows that the mass of the cluster decreases by $15\%$ at a time when the
number of stars has decreased by less than $11\%$.

These mean data actually mask a more complicated picture, which is given in
Fig.18.  At early times, i.e. during core collapse (Fig.18a) there is indeed
preferential escape of the stars of lowest mass, as might be expected from
H\'enon's theory and as a byproduct of mass segregation.  The actual fractional
escape is less than 1\% however.  After core collapse (Fig.18b) the
mass-dependence is reversed, as the most massive stars preferentially escape.
As
we have seen (Fig.3) there is little further mass segregation after core
collapse, and therefore part of the mechanism which led to the preferential
escape of the lightest stars is suppressed.  Now, however, interactions in the
core, which consists predominantly of stars of high mass, some in binaries,
lead
to escape of stars directly from the core, and this depletes the population of
stars of highest mass.

The number of binary escapers is very small. For example when $N=500$ the
average number is about $0.5$ at $t\sim 400,~\sim 20 t_{coll}$ (cf.  Table 2);
this number is about $1\%$ of all escapers (cf. Fig.17).  Simple theory implies
that the possibility of escape of binaries depends, among other things, on the
mass of stars which form the binary. The fraction of the kinetic energy which
the centre of mass of the binary (of total mass $m_b$) can gain during an
interaction with a field star of mass $m_3$ is approximately equal to
$m_3/(m_b+m_3)$. Therefore when $m_b > 2m_3$ this fraction is smaller than in
the case of equal masses.  It follows that a massive binary has to experience
more (or more energetic) interactions with field stars before its removal from
the system.  In a sense, therefore, it remains in the core longer than a binary
which consists of stars with masses equal to that of the single stars.

\medskip
{\bf 7.2 \enskip The energy budget}
\medskip

Initially the energy of the $N$-body systems consists entirely of the
``external" energy of the bound members, which we denote by $E_{ext}$.  When
binaries form, $E_{ext}$ increases (i.e. the cluster becomes less bound), and
the ``internal" energy of the bound binaries $E^b_{int}$ decreases in
compensation (Fig.15).  At the same time the escape of single stars by two-body
processes diverts energy into the external energy of the escapers
$E^{es}_{ext}$, and this tends to decrease the energy $E_{ext}^b$ of the
remaining bound stars.  Finally, as interactions with binaries become energetic
enough to lead to their escape (the energy associated with the motion of their
barycentres being denoted by $E^{eb}_{ext}$), the internal energy of escaping
binaries ($E^{eb}_{int}$) decreases (i.e. increases in magnitude), and the
internal energy of bound binaries $E^b_{int}$ increases (decreases in
magnitude)
in compensation.  The balance between these various processes is shown in
Fig.19.  Note that, although the number of binary escapers is small (\S7.1)
they
have a very significant effect on the energy budget.  It is worth comparing
Fig.19 with the corresponding diagrams published by Heggie \& Aarseth (1992)
for
larger systems ($N\simeq2500$) containing primordial binaries, though in Fig.19
we have split $E^e_{ext}$ into two components, corresponding, respectively, to
the kinetic energy of single stars and binaries.  Though the initial value of
$E^b_{int}$ is non-zero in that case, the trends in the data are remarkably
similar, even though the calculations of Heggie \& Aarseth were larger and used
equal masses.

\bigskip
{\bf 8 \enskip DISCUSSION AND CONCLUSIONS}
\medskip

We have analysed the evolution of isolated $N$-body systems in which the
spectrum of stellar masses is a power law.  The systems contained between 250
and 1000 stars, and for each $N$ we have averaged results from a large number
of
runs, in order to improve the statistical quality of the data.

Our results confirm the extreme rapidity of core collapse in the presence of a
mass spectrum (cf. Inagaki \& Wiyanto 1984).  Theoretically, the $N$-dependence
of the rate of core collapse involves the Coulomb logarithm $\ln\gamma N$, but
our results do not constrain the value of $\gamma$ as strictly as in the case
of
equal masses (Paper I).  Our results are consistent with a value around
$\gamma\simeq 0.02$, which is much smaller than the corresponding result for
equal masses, in qualitative agreement with the theory of H\'enon (1975).  For
such relatively small values of $N$ as in our simulations the interpretation of
the time scales is quite sensitive to the value of $\gamma$.  For example the
value of the initial half-mass relaxation time $t_{rh}(0)$ for $N=500$ varies
from about 20 for $\gamma=0.02$ to 11 for the standard value $\gamma=0.4$. If
the former value is adopted we find that core collapse is complete (Table 2) in
about $0.9t_{rh}(0)$.  Lest this seem much too short, it may be noted that
Chernoff \& Weinberg (1990), using a Fokker-Planck code, found that the time to
core collapse was also about $0.9t_{rh}(0)$ for the same mass spectrum, though
the initial model they adopted was a King model with dimensionless central
potential $W_0 = 3$, and a tidal cutoff was included.

One of our most curious findings (\S3.2 and Fig.3) is that the spatial
distribution of the mean stellar mass is nearly constant (i.e.
time-independent)
throughout the post-collapse evolution, when the overall expansion of the
system
is scaled out.  While this helps to explain why the time-dependence of the
post-collapse expansion takes the simple form of eq.(1), it is only an
empirical
explanation.  It remains to be understood on theoretical grounds why the
extremely rapid segregation of masses during core collapse comes to such an
abrupt halt.  One way in which this question might be approached would be to
try
to find a self-similar, multi-mass, post-collapse solution of the Fokker-Planck
or gas equations, in analogy with the equal-mass models of Goodman (1984, 1987)
or Inagaki
\& Lynden-Bell (1983).  It was suggested long ago by Lynden-Bell
(pers. comm.) that one might search for such a solution for core
collapse, in which the distribution of stellar masses was a power law.
Be that as it may, there is now empirical evidence that a search for
such a {\sl post}-collapse solution might be fruitful.

It is equally surprising, perhaps, that the departures from equipartition
(Fig.6) also follow a very simple behaviour in the post-collapse regime.
Certainly it may be expected that departures from energy equipartition will be
small in the core, where the relaxation time is short, and this is what is
found.  But it might have been expected that, at radii where the departure from
equipartition becomes significant, one should have observed a progressive
tendency towards equipartition.  However, we have found that there is
essentially no change in the profile of our equipartition parameter, except for
that caused by the nearly homologous expansion of the entire system.

Another surprise, except in hindsight, is the evolution of the heaviest stars.
We have seen (\S7.1) that they escape at a relatively high rate, which would
not
necessarily be expected because it might be thought that they would tend to
eject stars of low mass while settling, through mass segregation, into the
core.
Perhaps this does happen, but in the core the massive stars tend to form
energetic binaries, whose further interactions tend to eject these binaries and
the moderately massive stars in the core, thus enhancing their rate of escape.

Some hard binaries are ejected into the outer parts of the cluster without
completely escaping, and we noticed (\S6.2) that these tend to have higher
eccentricities than average, presumably as a result of the ejecting encounter.
Thus hard binaries beyond the half-mass radius tend to have large masses and
large eccentricities.  Though our models obviously differ in crucial respects
from the galactic globular clusters, it is possible to see here an example of
the dynamical behaviour which gives rise to such objects as the binary
millisecond pulsar in M15 (Phinney 1993).

\bigskip
\bigskip

{\bf ACKNOWLEDGMENTS}
\smallskip

{\parindent=0pt We would like to thank S.J. Aarseth for making available to us
a
version of his NBODY5 code, and S. Inagaki for the use of his Fokker-Planck
code. We are indebted to J.  Blair-Fish of Edinburgh Parallel Computer Centre
for much help in launching this program on the Edinburgh Computing Surface, a
transputer array operated by Edinburgh University Computing Service on behalf
of
the Edinburgh Parallel Computing Centre. This research has been supported by a
grant (number GR/G04820) of the U.K.  Science and Engineering Research
Council. }

\vfill\eject

{\bf REFERENCES}
\medskip
\leftskip=0.5 truein
\parindent=-0.5 truein
Aarseth S.J., 1985, in Brackbill J.U., Cohen B.I., eds,
Multiple Time Scales. Academic Press, New York, p.377

Aarseth S.J., Heggie D.C., 1992, MNRAS, 257, 513

Casertano S., Hut, P., 1985, ApJ, 298, 80

Chernoff D.F., Weinberg M.D., 1990, ApJ, 351, 121

Giersz M., Heggie D.C., 1994a, MNRAS, 268, 257 (Paper I)

Giersz M., Heggie D.C., 1994b, MNRAS, 270, 298 (Paper II)

Giersz M., Spurzem R., 1994, MNRAS, 269, 241

Goodman J., 1984, ApJ, 280, 298

Goodman J., 1987, ApJ, 313, 576

Gurevich L.E., Levin B.Yu., 1950, Astron. Zh., 27, 273 (also NASA TT F-11,541)

H\'enon M., 1969, A\&A 2, 151

H\'enon M., 1975, in Hayli A., ed, Dynamics of Stellar Systems, IAU Symp
69. Reidel, Dordrecht, p.133

Hills J.G., 1990, AJ, 99, 979

Inagaki S., Lynden-Bell D., 1983, MNRAS, 205, 913

Inagaki S., Wiyanto P., 1984, PASJ, 36, 391

Jeans J.H., 1929, Astronomy and Cosmogony, 2nd Edition. Cambridge
University Press, Cambridge, p.305

King I.R., 1966, AJ, 71, 64

Phinney E.S., 1993, in Djorgovski S., Meylan G., eds, Dynamics of Globular
Clusters. A.S.P. (Conf. Ser.), San Francisco, p.141.

Press W.H., Flannery B.P., Teukolsky S.A., Vetterling W.T., 1986, Numerical
Recipes. Cambridge Univ. Press, Cambridge.

Sigurdsson S., Phinney E.S., 1993, ApJ, 415, 631

Spitzer L., Jr., 1987, Dynamical Evolution of Globular Clusters.
Princeton Univ. Press, Princeton

\leftskip=0.0 truein
\parindent=0.0 truein

\vfill\eject

\centerline{\bf Table 1}
\medskip
\centerline{MASS GROUPS}
\bigskip
{\parindent=0pt {\SM
{}~~~range of masses ~~mean mass \hskip 4.4truecm initial data\hfill

\hskip 1.0truecm (units of $m_{min}$) \hskip 5.1truecm mass fraction and
number \hfill

\hskip 6.4truecm N=250 \hskip 2.5truecm N=500 \hskip 2.4truecm N=1000
$$\vbox{ \settabs \+ 1.00000 \enskip & \enskip 1.40390 \enskip & \enskip
1.01860
\enskip & \enskip 2.000000 \enskip & \enskip 2.000000 \enskip & \enskip
2.000000
\enskip & \enskip 2.000000 \enskip & \enskip 2.000000 \enskip & \enskip
2.000000
\enskip & \cr
\+\hfill1.000\hfill&\hfill1.439\hfill&\hfill1.186\hfill&\hfill0.1895\hfill&
\hfill101.1\hfill&\hfill0.1902\hfill&\hfill202.4\hfill&\hfill0.1987\hfill&
\hfill422.9\hfill &\cr
\+\hfill1.439\hfill&\hfill2.067\hfill&\hfill1.706\hfill&\hfill0.1722\hfill&
\hfill63.63\hfill&\hfill0.1705\hfill&\hfill126.1\hfill&\hfill0.1721\hfill&
\hfill254.3\hfill &\cr
\+\hfill2.067\hfill&\hfill2.970\hfill&\hfill2.454\hfill&\hfill0.1428\hfill&
\hfill36.71\hfill&\hfill0.1421\hfill&\hfill73.07\hfill&\hfill0.1440\hfill&
\hfill148.5\hfill &\cr
\+\hfill2.970\hfill&\hfill4.267\hfill&\hfill3.527\hfill&\hfill0.1161\hfill&
\hfill20.84\hfill&\hfill0.1198\hfill&\hfill42.88\hfill&\hfill0.1210\hfill&
\hfill86.55\hfill &\cr
\+\hfill4.267\hfill&\hfill6.132\hfill&\hfill5.054\hfill&\hfill0.09463\hfill&
\hfill11.85\hfill&\hfill0.09874\hfill&\hfill24.66\hfill&\hfill0.09822\hfill&
\hfill49.03\hfill &\cr
\+\hfill6.132\hfill&\hfill8.810\hfill&\hfill7.213\hfill&\hfill0.08564\hfill&
\hfill7.446\hfill&\hfill0.07981\hfill&\hfill13.96\hfill&\hfill0.08138\hfill&
\hfill28.35\hfill &\cr
\+\hfill8.810\hfill&\hfill12.66\hfill&\hfill10.34\hfill&\hfill0.06648\hfill&
\hfill4.054\hfill&\hfill0.06978\hfill&\hfill8.518\hfill&\hfill0.07357\hfill&
\hfill17.85\hfill &\cr
\+\hfill12.66\hfill&\hfill18.19\hfill&\hfill15.01\hfill&\hfill0.06046\hfill&
\hfill2.554\hfill&\hfill0.05607\hfill&\hfill4.714\hfill&\hfill0.05732\hfill&
\hfill9.675\hfill &\cr
\+\hfill18.19\hfill&\hfill26.13\hfill&\hfill21.36\hfill&\hfill0.04526\hfill&
\hfill1.321\hfill&\hfill0.04412\hfill&\hfill2.607\hfill&\hfill0.04220\hfill&
\hfill4.975\hfill &\cr
\+\hfill26.13\hfill&\hfill37.50\hfill&\hfill31.18\hfill&\hfill0.02588\hfill&
\hfill0.5357\hfill&\hfill0.02777\hfill&\hfill1.125\hfill&\hfill0.03547\hfill&
\hfill2.900\hfill &\cr}$$
}}

\smallskip
Note: the mass fractions and numbers were determined from the sets of
initial conditions.  For an analytical mass function, the mass
fraction would be independent of $N$, and the number would be
proportional to $N$.

\vfill\eject

\centerline{\bf Table 2}
\medskip
\centerline{SUMMARY OF RESULTS}
$$\vbox{ \settabs \+\quad 111111 \quad & \quad 111111111 \quad & \quad 111111
\quad & \quad 111111 \quad & \cr
\+ \hfill $N$ \hfill & \hfill $t_{coll}$ \hfill & \hfill $r_c/r_h$ \hfill &
\hfill $N_c$ \hfill &\cr
\smallskip
\+ \hfill 250 \hfill & \hfill 12 ~(95) \hfill & \hfill 0.11 \hfill &
 \hfill 4 \hfill & \cr
\+ \hfill 500 \hfill & \hfill 18 ~(164) \hfill & \hfill 0.14 \hfill & \hfill 8
 \hfill & \cr
\+ \hfill 1000 \hfill & \hfill 21 ~(332) \hfill & \hfill 0.16 \hfill &
\hfill 17 \hfill & \cr}$$
\medskip
\noindent Note:
$t_{coll}$ is the approximate time of the end of core collapse (determined by
the minimum of the central potential, cf. fig.12).  Values for equal masses
(from the data discussed in Papers I and II) are shown in brackets.  The values
of $r_c/r_h$ and $N_c$ are representative of almost all of the post-collapse
evolution.

\centerline{\bf Table 3}
\medskip
\centerline{FITTED PARAMETERS}
\medskip

$$\vbox{ \settabs \+ ~~10000~~ & ~~0.08200~~ & ~~1000000~~ & ~~1000000~~ &
{}~~1000000~~ & ~~1000000~~ &\cr
\+ \hfill $N$ \hfill & \hfill $t_0$ \hfill & \hfill $a$ \hfill & \hfill $\nu$
 \hfill & \hfill $b$ \hfill & \hfill $c$ \hfill & \cr
\smallskip
\+ \hfill 250 \hfill & \hfill  -5.688 \hfill & \hfill 316.94 \hfill &
\hfill -0.067 \hfill & \hfill 0.159 \hfill & \hfill 16.95 \hfill &\cr
\+ \hfill 500 \hfill &  \hfill -8.256 \hfill & \hfill 622.95 \hfill &
\hfill -0.063 \hfill & \hfill 0.095 \hfill &\hfill 17.24 \hfill &\cr
\+ \hfill 1000 \hfill & \hfill +4.661 \hfill & \hfill 1236.6 \hfill &
\hfill -0.061  \hfill&  \hfill 0.064 \hfill &\hfill 17.25 \hfill & \cr}$$
\medskip
Note: the symbols are explained in eqs.(1) and the accompanying discussion.
\vfill\eject

{\bf FIGURE CAPTIONS}
\bigskip
{\bf Fig.1}  Evolution of {\sl projected} Lagrangian radii for $N = 500$.  Also
shown is the (three-dimensional) core radius, defined in \S5.

{\bf Fig.2} Evolution of the projected (surface) mass density profile for $N =
500$.  The radial coordinate is scaled by the instantaneous projected half-mass
radius.  The densities plotted are average values between each Lagrangian
radius, and have been plotted at the mean radius in each Lagrangian shell.

{\bf Fig.3} Evolution of the mean individual stellar mass within each projected
Lagrangian shell for $N = 500$.  Thus the curve labeled ``$2\%$'' gives the
mean
masses of stars whose projected distance from the density centre lies between
projected radii which enclose $1\%$ and $2\%$ of the projected mass,
respectively.  A binary was treated as a single star with a mass equal to the
total mass of the components.

{\bf Fig.4}  Profile of projected mass density of each mass class at time
$t=300$ for $N = 500$. The Lagrangian radii have been determined separately for
each mass class, and since the plotted values of $\Sigma$ represent
average estimates for annuli separated by these projected Lagrangian
radii, the innermost and outermost plotted points vary from one species
to another.

{\bf Fig.5}  Root mean square projected  velocity in each projected
Lagrangian annulus for $N = 1000$.  The data have been cleaned by removal of
energetic escaping particles, which would otherwise cause a ``spike",
especially in the outer shells, where they spend the longest time.  Also
plotted ($v_c$) is the rms projected velocity within a {\sl
sphere} whose radius equals the
core radius. (Note that this quantity is {\sl not} a projected
velocity dispersion in the usual sense.)

{\bf Fig.6} Projected equipartition parameter $B$ (eq.(2)) at various times in
a
set of models with $N = 500$.  The radial coordinate has been scaled by the
projected half-mass radius.

{\bf Fig.7} Projected velocity dispersion profiles for each mass group for $N =
500$ at time $t =300$.

{\bf Fig.8} Projected anisotropy profiles for $N=500$.

{\bf Fig.9}  Projected anisotropy profiles at $t=300$ for each mass class in
models with $N = 500$.

{\bf Fig.10} Mass-weighted root mean square central speed as a function of
time.  The result for $N = 250$ is plotted to scale, while those for $N
= 500$ and $1000$ have been displaced vertically by $0.3$ and $0.6$
units, respectively.  The data have been smoothed over two time units.

{\bf Fig.11} The projected mass density in the eight mass groups within the
innermost Lagrangian shell (i.e. the 1\% Lagrangian radius), for $N=500$.

{\bf Fig.12} Evolution of the central potential (i.e. at the
location of the density centre) for $N = 250,~500$ and
$1000$.

{\bf Fig.13} Evolution of core parameters $\rho_c$ and $v_c$ for $N = 500$.
The straight line corresponds to $\rho_c\propto v_c^6$. The $N$-body data
were smoothed using the standard routine SMOOFT from Press \et  (1986).

{\bf Fig.14}  Number of hard binaries as a function of time, for $N = 250,~500$
and $1000$.  Only binaries which are not regarded as escapers are included.

{\bf Fig.15} Total internal energy of all bound hard binaries for $N = 250$,
$500$ and $1000$.

{\bf Fig.16}  Mean mass of hard binaries in 4 spatial zones for $N = 500$.
$M_{tot}$ is the total bound mass of the system.

{\bf Fig.17} Evolution of total number and mass of bound stars in systems with
$N = 500$ stars initially, expressed as a fraction of the initial value.

{\bf Fig.18}  Fractional escape of stars in different mass classes, for $N =
1000$. Fig.18a shows details for the collapse phase and the start of the
post-collapse evolution, while Fig.18b shows the long-term trend.

{\bf Fig.19}  Energy balance for $N = 1000$.  The meaning of the symbols is
given in the text.
\bye